\documentclass[mathleft
]{an}
\usepackage{graphicx}
\usepackage{times}
\usepackage{natbib}
\def\aj{AJ}%
\def\apj{ApJ}%
\def\apjl{ApJ}%
\def\apjs{ApJS}%
\def\aap{A\&A}%
\def\aapr{A\&A~Rev.}%
\def\mnras{MNRAS}%
\def\nat{Nature}%

\begin{document}

\Pagespan{1}{}
\Yearpublication{2006}%
\Yearsubmission{2005}%
\Month{11}%
\Volume{999}%
\Issue{88}%

\title{Early-type dwarf galaxies in clusters: a mixed bag with various origins?}

\author{Thorsten Lisker\thanks{Corresponding author:
  \email{TL@x-astro.net}\newline}
}
\titlerunning{Early-type dwarf galaxies}
\authorrunning{Thorsten Lisker}
\institute{
Astronomisches Rechen-Institut, Zentrum f\"ur Astronomie der
  Universit\"at Heidelberg, M\"onchhofstra\ss e 12-14, 69120
  \mbox{Heidelberg}, Germany}

\received{}
\accepted{}
\publonline{}


\abstract{%
The formation of early-type dwarf (dE) galaxies, the most numerous objects in clusters, is believed
to be closely connected to the 
physical processes that drive galaxy cluster evolution, like galaxy
harassment and ram-pressure stripping. However, the actual
significance of each mechanism for building the observed cluster dE
population is yet unknown.
Several distinct dE subclasses were
 identified, which show significant differences in their shape,
stellar content, and distribution within the cluster. Does this
diversity imply that dEs originate from
various formation channels? Does ``cosmological'' formation play a
role as well? I try to touch on these
questions in this brief overview of dEs in
galaxy clusters.
%
}

\maketitle

\section{Introduction}

\begin{figure}
\centering
\includegraphics[width=50mm]{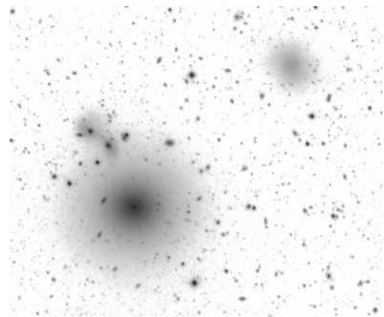}
\caption{Two Virgo cluster dEs (VCC\,856 \& 839) in a deep exposure with ESO 2.2m/WFI. The image
  shows an area of $6'\times5'$.
}
\label{dEbigsmallfig}
\end{figure}

Early-type dwarf (dE) galaxies play a key role in understanding galaxy
cluster evolution. Their low mass and low density make them more
susceptible to physical effects than giant galaxies, and they are a
large population, outnumbering all other galaxy types in dense
environments by far. They are thus ideal probes of the mechanisms that
govern galaxy formation and evolution in a cluster environment. Nevertheless,
dEs are also interesting in their own right. Despite their rather
unspectacular appearance, a surprising complexity in their characteristics has
become evident, in terms of kinematics (rotating vs.\ pressure
supported), structure (flat vs.\ round), and spatial distribution
(cluster center vs.\ outskirts). This zoo of dEs and their possible
origin(s) is an ongoing challenge for observers and theorists.



\section{Morphology and classification}

\subsection{What is an early-type dwarf?}

\begin{figure}
\centering
\includegraphics[width=72mm]{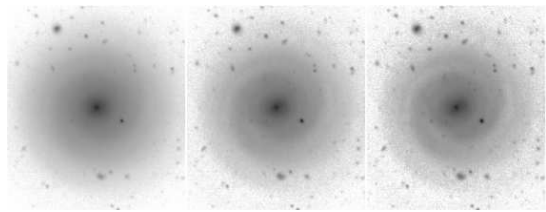}
\caption{The galaxy VCC\,0308, classified as dwarf S0 by \citet{vcc},
  in a deep exposure with ESO 2.2m/WFI (left). A residual image of the
  weak spiral arm structure was created by subtracting a model of
  the smooth light distribution. This residual image was then
  multiplied by a factor of 2 and 3, and added again to the smooth
  model, yielding the middle and right images.}
\label{dEspiral}
\end{figure}

Morphological classification of {\it early-type} galaxies can be
subjective to a certain extent, and the division between {\it dwarf}
and normal/giant galaxies is often rather arbitrary. It thus appears
reasonable to consider these two terms separately.

\paragraph{Early-type:}
Elliptical (E) and lenticular (S0) galaxies are commonly subsumed
under the term {\it early-type} galaxies. Their main characteristic
is the absence of substructure like spiral arms or strong asymmetries:
their overall appearance is smooth and regular
(Fig.~\ref{dEbigsmallfig}).
A complication is added by the fact that
weak spiral structure has been identified in a number of early-type
dwarfs by image processing techniques. In some cases, the spiral
features extend across the whole galaxy
(Fig.~\ref{dEspiral}). Should such an object be re-classified as
spiral galaxy, once the spiral arms are discovered? What if the spiral
arms are strong enough in one galaxy to be seen in the initial image,
but not in another galaxy, for which special techniques are necessary?
B.\ Binggeli identified a handful of ``dwarfish looking S0/Sa''
galaxies in the Virgo cluster \citep[unpublished, see][]{p1}, whose
appearance is similar to the illustration in
Fig.~\ref{dEspiral}. Here, classification (and the subsequent
selection of a working sample!) becomes arbitrary
to some extent, and it would be more important to locate those galaxies in
their physical parameter space.



\paragraph{Dwarf:}
While the term {\it dwarf} would imply that an object is small and faint,
Figure~\ref{dEbigsmallfig} illustrates that a brighter dwarf might appear as an object of
rather high surface brightness when compared to fainter, ``low surface
brightness'' dwarfs \citep[cf.][]{Sabatini2003,Adami2006}, since 
size and surface brightness are correlated with total magnitude\linebreak
\citep{bin91}.
{\it Dwarf elliptical} is a widely used term,
indicating that the overall appearance is similar to that of an
E galaxy, yet reduced to dwarf scale. However,
another class of elliptical galaxies also has dwarf
sizes and magnitudes, but a surface brightness similar to
the giants: the {\it compact ellipticals} or {\it
  low-luminosity ellipticals} \citep{deV61}, with M32 typically considered as
prototype \citep[but see][]{GrahamM32}.

Some authors \citep{Moore1998,Kormendy2009} instead refer to
early-type dwarfs as {\it
  spheroidals} (Sph). This avoids implying a
priori a physical relation between Es and early-type dwarfs, since the
latter could also be more closely related to late-type galaxies. However, 
{\it spheroidal} is a threedimensional description, which can be
somewhat impractical for astronomical classification, since we only
see the projected twodimensional appearance. 
Many ``spheroidals'' are (now) known to
be of relatively flat, disk-like shape \citep{p3}. In
contrast, the term {\it elliptical} applies, to first approximation,
even to the disky early-type dwarfs, since they
still have a nearly elliptical projected shape.

\begin{figure}
\includegraphics[width=82mm]{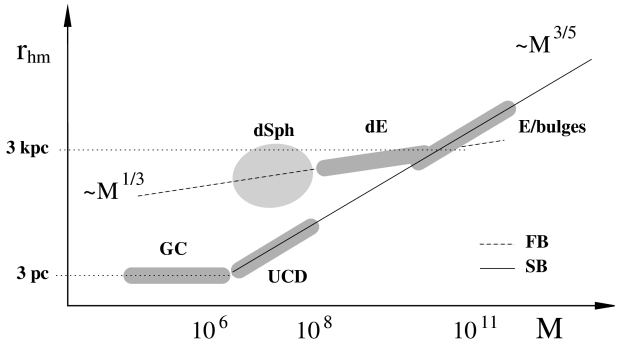}
\caption{Sketch of the relation of half-mass radius and
  dynamical mass (in ${\rm M_\odot}$) for dynamically
  hot stellar systems (from \citealt{Pflamm2009}, based on \citealt{Dabringhausen2008}).}
\label{dabfig}
\end{figure}

Using surface brightness as the main criterion to separate dEs from Es
\citep{san84}, and thereby creating a significant overlap of their
total magnitudes\linebreak \citep{Jerjen1997}, might indeed be a more
``physical'' way to go: early-type galaxies seem to arrange themselves
in two main sequences (Fig.~\ref{dabfig}). One sequence follows the
more diffuse objects, from dEs to the fainter dwarf spheroidals
(dSphs), the other sequence contains the more compact or centrally
concentrated objects, ranging from massive Es to spiral bulges
\citep{Graham2008} and compact ellipticals \citep{Kormendy2009},
probably continuing to ultra-compact dwarf galaxies and the brightest
globular clusters \citep{Dabringhausen2008}.
Given these relations, it might be the clearest (although not the most
common) expression to call
the dEs {\it diffuse ellipticals} \citep{Prugniel1996}, or more
general {\it \mbox{diffuse early-type galaxies}}.

\subsection{Structural characteristics}
\label{sec:sub_structure}

\begin{figure}
\includegraphics[width=82mm]{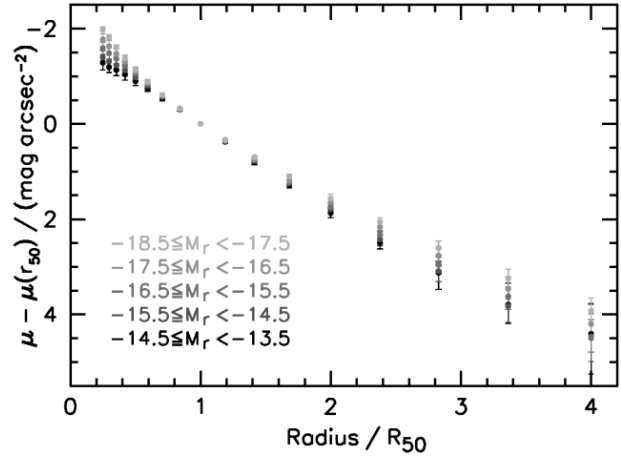}
\caption{Relative radial surface brightness profiles of Virgo dEs in SDSS-$r$,
  calculated at each position as the median value of all Virgo cluster
  member dEs in a given magnitude interval. The error bars indicate
  the range from 25\% to 75\% of the values.}
\label{profilefig}
\end{figure}

Early-type dwarfs have surface brightness (SB) profiles less steep than
massive ellipticals -- yet the frequently quoted  ``exponential vs.\ de
Vaucouleurs'' scenario is too simplistic. In fact, SB profiles of
both dwarf and giant early types can be
well described by continuous S\'ersic profiles \citep{Sersic1963}, with the
logarithm of the S\'ersic index $n$ linearly increasing with central
surface brightness and magnitude
\citep{Young1994,Graham2003a,Gavazzi2005a}. This is illustrated for Virgo dEs in
Fig.~\ref{profilefig}, where an exponential profile would be a
straight line, which clearly does not fit the brighter dEs.

While a clear correlation exists between (effective or central) surface brightness
and magnitude \citep{bin91}, albeit with significant scatter
\citep{Binggeli1998}, the size-magnitude correlation is somewhat less
well defined (also see Janz \& Lisker, this issue), and the half-light
radius decreases only slightly towards fainter galaxies
\citep{Misgeld2009}, with typical values around 1\,kpc
\citep{SmithCastelli2008}.
Whether dEs and Es form a continuous family in all basic scaling
relations is still debated (compare, e.g., \citealt{Graham2003a} and
\citealt{Kormendy2009}).
While the correlation between S\'ersic index $n$ and magnitude is not
put into question, \citet{Janz2008} showed that, at those magnitudes
where dEs and Es approach each other and eventually overlap, most dEs
are larger and most Es are smaller than the expected one-family relation for
varying $n$ \citep[also compare][]{Misgeld2009}. However, no significant systematic departure
from this relation was found by \citet{Graham2008}.
\citet{Kormendy2009} interpret the different behaviour of Es and dEs
with fundamentally different formation mechanisms.

\begin{figure}
\includegraphics[width=82mm]{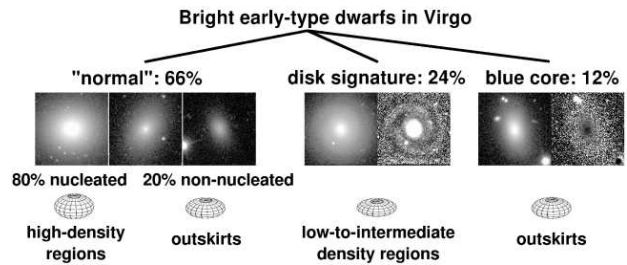}
\caption{Scheme of dE subclasses, adapted from \citet{p3}. Only bright
  Virgo dEs ($M_r<-16\fm0$, using $m-M=31\fm0$) are considered. The
  small ellipsoid drawings are based on the median of the estimated intrinsic
  axial ratios for each subclass.}
\label{treefig}
\end{figure}

Analogous to the bright S0s, \citet{san84} introduced the class of
dwarf S0 (dS0), defined by having the overall appearance of a dE, but
exhibiting features that probably indicate the presence of a
disk. These criteria were, a bulge-disk-like structure, high
flattening, a lens-like surface brightness distribution, a global
asymmetry like boxiness or a bar, or a central irregularity\linebreak \citep{bin91}.
Later, weak spiral structure was identified in a dE for the first time
\citep{Jerjen2000a}, with similar discoveries following
\citep[e.g.][]{Barazza2002a,Graham2003b}. A systematic search for weak
disk signatures was carried out by \citet{p1} for Virgo cluster dEs,
revealing such features (Fig.~\ref{dEspiral}) in one quarter of the brighter dEs
(Fig.~\ref{treefig}), with the ``disky fraction'' even reaching 50\%
at the bright end. While these identifications do have a 
large overlap with the dS0 classification in the Virgo cluster catalogue
\citep[VCC,][]{vcc}, a significant number of dS0s remained where no
disk features were found. These were mainly those with blue central
colours from young stars \citep{p2}, where central irregularities and
dust structures are seen. Therefore, \citet{p3} suggested a new
subclassification, starting with {\it dE} (for dwarf or diffuse
early-type galaxy) and adding in brackets {\it (di)} for
disk, {\it (bc)} for blue center, {\it (N)} for nucleated (which had
been {\it dE,N} in the VCC), and {\it (nN)} for non-nucleated. The
latter was to avoid the ambiguity that {\it dE} could either refer to
dEs in general, or explicitly to non-nucleated dEs.  Nevertheless,
\citeauthor{p3} were somewhat inconsequent, in that they did not use
this subclassification in a cumulative way, e.g.\ dE(N,di) for a
nucleated dE with disk substructure, or dE(nN,di,bc) for a
non-nucleated dE with disk substructure and blue central
colors. Furthermore, having a blue center is of course no
\emph{morphological} feature.

A different approach to identify dS0 galaxies, based on the
two-component criterion of \citet{bin91}, was applied to Coma cluster
dEs by \citet{Aguerri2005}: those objects whose SB profile was not
well fitted by a single S\'ersic profile, but required a
S\'ersic+exponential fit, were classified as dS0s. Like
\citet{Binggeli1995} for the Virgo cluster, \citet{Aguerri2005} found
the dS0s to be flatter than the ordinary dEs. Furthermore,\linebreak
\citet{Ferguson1989} and \citet{Ryden1994} showed that the non-nucleated dEs are flatter than
the nucleated ones. From the distributions of projected axial ratios,
\citet{p3} estimated intrinsic axial ratio distributions for the
different subclasses, which are illustrated in Fig.~\ref{treefig}:
indeed, dE(nN)s are flatter than dE(N)s (the difference
disappears towards fainter magnitudes), dE(bc)s have similarly flat
shapes as the dE(nN)s, and dE(di)s are the flattest
population.


The classification of a dE as nucleated
  and non-nucleated is not unambiguous. Many 
  Virgo dEs that were classified as non-nucleated in the VCC actually
  host a faint nucleus that was hardly detectable with the VCC data
\citep{Grant2005,Cote2006}.
A more appropriate term might be \emph{dEs without a nucleus of
  significant relative brightness} as compared to the central galaxy
light. \citet{Grant2005}
pointed out that dEs classified as nucleated and non-nucleated might
actually form a continuum with respect to relative nucleus
brightness. This makes the identification of systematic differences
between them even more interesting. For the sake of simplicity, I
continue to use {\it non-nucleated} for those without a
``significant'' nucleus.
Due to the limited space, I do not further discuss the central
structural properties of dEs \citep[e.g.][]{Cote2007}, their nuclei
\citep[e.g.][]{Lotz2004}, and their globular cluster
systems \citep[e.g.][]{Peng2006,Peng2008}.

\section{Colours and stellar populations}

\subsection{Age and metallicity}

Studies of Lick absorption line indices in galaxy spectra revealed
intermediate to old ages ($\sim5$\,Gyr, with rather large scatter),
subsolar metallicities ($-1 \la [Fe/H] \la 0$), and more or less solar 
$\alpha$-element abundances
\citep[e.g.][]{Geha2003,vanZee2004b,Michielsen2008,Koleva2009,Chilingarian2009,Paudel2009}.
From narrowband photometry, \citet{Rakos2004} inferred an age bimodality:
they found that nucleated dEs are several Gyr older than non-nucleated
dEs, and have old ages comparable to globular clusters.

Stellar population gradients were studied for the brighter dEs by
\citet{Chilingarian2009} and \citet{Koleva2009}, who found a range of metallicity gradients
between steeply radially decreasing and constant metallicity, but no
positive gradients. The stellar population age remains constant or
shows a slight increase with radius. These findings would seem consistent
with the simple picture that all dEs had their last star formation
activity in their centers, as seen in those dEs with blue cores: if
gas could be retained in the central region, while being stripped much
earlier in the outskirts, the metal enrichment would be higher in the
center, and the average age would be younger due to the longer
duration of star formation there. Note also that the nuclei or nuclear
regions were found to have younger ages than their host galaxies (Paudel
\& Lisker, this issue; \citealt{Chilingarian2009}).

While detailed studies of resolved stellar populations of cluster dEs
are hardly possible at distances of Virgo and beyond, the first steps
were already done: \citet{Caldwell2006} and \cite{Durrell2007}
presented analyses of resolved red giant branch populations for dEs
and a dSph in the Virgo cluster.

\subsection{Relations with luminosity}

A fundamental characteristic of early-type galaxies is the
well-defined relation between colour and magnitude, with more luminous
objects having redder colours 
\citep[e.g.][]{sanvis78a,cal83}.
While this trend includes
the dEs, there has been disagreement about whether or not they follow
the same CMR as the giant ellipticals, and whether this CMR is linear.
Although \citet{cal83} reported a linear CMR over a range of
$-15 \le M_V \le -23$\,mag, his Figure~3 actually seems to indicate
that the CMR might slightly bend towards the dEs, such
that with decreasing magnitude, dwarfs become bluer more
rapidly than giants do. This is similar to the non-linear CMR of Virgo
cluster dwarf and giant early types that was reported by
\citet{Ferrarese06}, who found that a parabolic curve provided a good
fit to the data. However, \citet{Andreon2006} remarked that, when
taking into account an intrisic scatter in the statistical 
analysis, the data of \citeauthor{Ferrarese06} do not support this
parabolic model, and are consistent with a linear
relation. \citet{Andreon2006} themselves report a linear CMR in the
cluster Abell 1185.

Interestingly, the conclusion of \citet{deV61},
that the dwarfs are ``systematically bluer'' than the giants, was
confirmed by \citet{Janz2009a}, who used SDSS $ugriz$ multiband photometry
for several hundred Virgo cluster early-type galaxies, and found a CMR
with an S-like shape (see Janz \& Lisker, this issue). This is caused
by an apparent transition region at intermediate magnitudes, where
the CMR slope is different than brightward and faintward of it.
While the CMRs presented by \citet{Misgeld2008,Misgeld2009} for the
Hydra I and Centaurus clusters appear still consistent with the Virgo results,
they are well fitted with a linear CMR. A linear relation from
dwarf to giant early types was also found in the Antlia cluster by\linebreak
\citet{SmithCastelli2008}.

Two results are common to all of those studies: i) there is no gap
between dwarfs and giants, and ii) the CMR is still well defined even
at faint ($M_B\approx-14$\,mag) magnitudes, pointed out by
\citet{p4} and \citet{SmithCastelli2008}.
The CMR of dEs in the Perseus cluster presented by
\citet{conIII} showed a considerable increase in scatter at magnitudes
$M_B \ge -15$\,mag, apparently caused by a colour bimodality of faint
dwarfs. However, this result was partly revised by \citet{Penny2008},
who showed with new spectroscopic measurements that most of the
redder objects are not members of the
cluster. Still, a significant increase in the scatter remains, caused
by objects bluer than the CMR extrapolation.

A further investigation of the CMR is possible by dividing the dEs
into different subsamples. \cite{p4} found that Virgo dEs sitting in
projected cluster regions of different local galaxy density exhibit a
slightly different CMR slope: those in higher-density regions are
redder at brighter magnitudes as compared to dEs in lower-density
regions, while their colours are similar at fainter
magnitudes. The same is true when comparing nucleated with
non-nucleated
dEs: the nucleated ones are redder at brighter magnitudes
\citep{p4}, consistent with the conclusion
 that they have older ages \citep{Rakos2004}.

\citet{fab73} showed that the strength of absorption features in the
spectra of galaxies is related to their luminosity. This can be seen
as the spectroscopic analogue to the CMR. \citeauthor{fab73}
interpreted the CMR as a trend of increasing metallicity with
luminosity, which today is still considered to be the primary
determinant of the CMR \citep[e.g.][]{kod97,cha06}. This is
illustrated, for example, in \citet{Geha2003}, who presented a clear
correlation of the spectral Mgb and $<$Fe$>$ absorption indices with
galaxy velocity dispersion from dwarfs to giants, with higher
metallicities at higher masses \citep[see also][]{Michielsen2008}. The
correlation between colour and metallicity $[Fe/H]$ was directly shown
by \citet{held1994}. 

Nevertheless, there is also evidence that the average stellar
population age decreases with decreasing luminosity from early-type
giants to dwarfs. In a multiwavelength
spectrophotometric study, extending from the ultraviolet (UV) to the
near-infrared (NIR), \citet{gavazzi2002} found that the timescale
for the duration of star formation increases with decreasing galaxy
luminosity, when fitting galaxy SEDs to population
synthesis models with fixed formation epoch. Similarly, the $H\beta$
absorption index, which serves as indicator for recent star formation
and stellar age, was found to increase with decreasing luminosity
\citep{Poggianti2001} and velocity dispersion
\citep{Geha2003} from giants to dwarfs. Thus the CMR could, at least
partly, also be governed by the stellar population age. Note, however,
that the above-mentioned correlations were mainly defined by comparing
giant early types and brighter dwarfs. While indeed the former have
higher stellar population ages than the latter \citep{Michielsen2008},
it is not yet clear whether and how age and luminosity are correlated
\emph{within} the dwarf regime. A large range of dE ages was found by
\citet{Michielsen2008}, and there are even indications for an
anticorrelation of age and luminosity  \citep{Paudel2009}.

The UV photometry of GALEX \citep{GALEX} recently brought new insight
into these issues. \citet{Boselli2005} presented the near-UV (NUV) and
far-UV (FUV) colours of Virgo early-type galaxies. The CMR is again
continous from dwarfs to giants in UV$-$optical and UV$-$NIR
colours. However, while its shape appears rather linear in
NUV$-V$ and NUV$-H$, except for the region of constant colour at the 
highest luminosities \citep[cf.][]{Janz2009a}, a clear bend is present
when NUV is replaced by FUV, or even stronger when FUV$-$NUV is
considered. In the latter case, while dwarfs become bluer with decreasing
luminosity, giants do so with \emph{increasing} luminosity, probably
explained by the UV-upturn phenomenon \citep{GildePaz2007}.
There might even be a break between
the dwarf and giant parts of the CMR, but a definite conclusion is
hampered by sparse sampling at intermediate luminosities.
 Since this
wavelength region is governed not only by the intermediate-age stars
responsible for the UV-upturn, but of course also by very young stars,
the interpretation of \citet{Boselli2005} was that residual star
formation activity is present in the dEs. However, based on the
above-mentioned idea of continuously decreasing stellar age with
decreasing galaxy luminosity, \citet{LiskerHan2008} remarked that the
observed colour behaviour would be consistent with the expectations from
population synthesis models if a continous increase in the duration of star
formation from giants to dwarfs was assumed. The real situation might
be more complex: as shown by Kim et al.\ (this issue), significant
differences become apparent when analysing the UV$-$optical colours
separately for dEs with and without disks, with the former being
systematically bluer, and for dEs in the center and outskirts of the
cluster, with the latter clearly having bluer colours.

\section{Kinematics}


After it has become clear that at least some early-type dwarfs do show
significant rotation \citep{BenderNieto1990}, the
questions were (and still are), how many of them rotate, and how strong is
the rotation? This was addressed by several studies through
medium-resolution spectroscopy in one or two dimensions
\citep{Simien2002,Pedraz2002,Geha2002,vanZee2004a,DeRijcke2005,Chilingarian2009},
finding significant rotation in some, but not all dEs. In many cases,
 the rotation curve is still rising at the end of
the radial extent of the data.

The Faber-Jackson relation \citep{FaberJackson} of galaxy luminosity
and velocity dispersion clearly flattens when
going from luminous Es to dEs \citep{DeRijcke2005,mat05}. The
Tully-Fisher relation\linebreak \citep{TullyFisher} of spiral and irregular
galaxies was shown by \citet{vanZee2004a} to be followed by dEs with
significant rotation. 
Note, however, that if these systems were pure disks, their
relatively high velocity dispersions would cause a large asymmetric
drift effect on the measured 
velocity \citep{DeRijcke2007,LiskerFuchs2009}. Dynamical models are
thus used by \citet{DeRijcke2007} to infer the circular velocities,
finding that dEs fall slightly below the baryonic Tully-Fisher
relation followed by early-type and late-type galaxies. Furthermore,
 at least for a single Virgo dE with weak spiral
structure (VCC\,856), \citet{LiskerFuchs2009} concluded that the
galaxy most likely has a significant dynamically hot component, in
which a disk with much lower velocity dispersion is embedded, thus
having only a small asymmetric drift effect.

 Common practice
is to compare the measurements of rotational velocity $v$ and velocity
dispersion $\sigma$ with the theoretical curve for an
isotropic oblate spheroid flattened by rotation and viewed edge-on
\citep{Binney1978}, as exemplified in \citet{Simien2002}. This curve can be
approximated analytically, following \citet{Kormendy1982proc}, so that
the ratio $v/\sigma$ can be normalized with respect to the curve:
\begin{equation}
(v/\sigma)^* = \frac{v/\sigma}{\sqrt{\epsilon/(1-\epsilon)}}\quad,
\end{equation}
with $\epsilon$ being the measured (projected)
ellipticity. $(v/\sigma)^*$ is frequently called the anisotropy
parameter, since in case of $(v/\sigma)^*\ll 1$ and a non-spherical
shape of the galaxy, it indicates that the flattened shape must be
caused by anisotropic velocity dispersion.

The terms used to interpret the values of $(v/\sigma)^*$, like {\it
  rotationally supported} or {\it pressure-supported}, are not always
applied in the same way or with the same criteria.
In particular, it is not always clear whether {\it rotational support} is
meant as support against gravity, or just as support of
  the galaxy's shape, i.e.\ being {\it rotationally flattened}. For
  example, a close-to-spherical galaxy with just a small ellipticity
  needs   only a little amount of rotation to make it above the
  theoretical line. Clearly, this galaxy would not be rotationally
  supported against gravity, but would still be pressure supported,
  albeit with some rotational flattening. On the other hand, we know
  for those galaxies where weak spiral arms or bars were identified,
  that they are not seen edge-on, and that they are relatively flat
  intrinsically --- it is thus not
  straightforward to speak of pressure or rotational support just on
  the basis of $(v/\sigma)^*$.

  The current inventory of rotation curves for dEs is, unfortunately,
  not very impressive -- owing to the fact that these faint galaxies
  require much longer exposure times than their giant
  counterparts. For the Virgo cluster, I collected published tabulated
  values of $v$ and $\sigma$ (mostly $\sigma_0$) from the studies
  named above, judged the radial
  extent of the rotation curve visually from the published rotation
  curve figures (using the second to last data point), compared it to
  the half-light semimajor axis ($a_{\rm hl}$) in SDSS-$r$ from \citet{p3}, and
  excluded those not reaching $1/3 a_{\rm hl}$. For galaxies appearing
  in more than one publication, I took those values that correspond to
  the largest radial extent, even if the quoted errors were
  larger in this reference as compared to others. This results in only
  27 dEs with useful values, 11 of them (41\%) having $(v/\sigma)^*\ga 1$,
  i.e.\ being flattened by rotation. As discussed above, a value for
  being {\it rotationally supported} might be rather arbitrary; when choosing
  $(v/\sigma)^*\ga 1.3$, 5 objects (19\%) remain. Interestingly,
  the correlation with those dEs with identified disk features
  \citep{p1} is not (yet) significant: of the 12 dEs with disks
  among the 27 galaxies (44\%), only 6 (55\%) belong to the group with
  $(v/\sigma)^*\ga 1$.

\begin{figure}
\includegraphics[width=76mm]{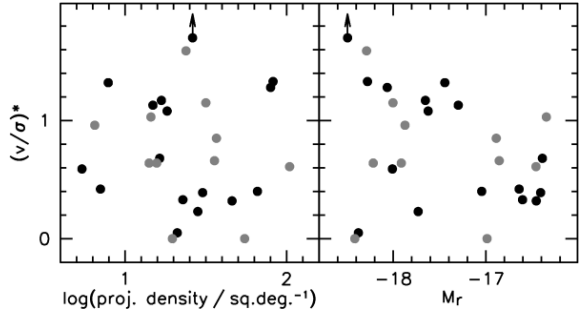}
\caption{Anisotropy parameter versus local projected galaxy density
  (left) and absolute magnitude (right) for Virgo
  dEs (see text). Grey symbols represent
rotation curves not reaching $0.8 a_{\rm hl}$.}
\label{rotfig}
\end{figure}

  Of those 27 galaxies, only 16 reach beyond $0.8 a_{\rm hl}$ in their
  rotation curve extent, and only 11 actually reach the point of
  half light. Compared with the number of Virgo dEs in the same magnitude
  range ($M_r<-16.3$, using $m-M=31$), 101 galaxies, this is a rather
  small fraction of only 11\%. Nevertheless, it is sufficient for taking
  a look at possible relations with environmental density and galaxy
  magnitude (Fig.~\ref{rotfig}). No robust correlation with local
  projected density can be
  claimed, but at least a trend is seen that, among the objects
  whose curves do reach the half-light radius, those that are clearly
  pressure supported are located in the high-density
  regions. \citet{Toloba2009}, who extend the sample of dEs with
  kinematical measurements by their new data, find a clearer relation
  when plotting $(v/\sigma)^*$ against clustercentric radius
  \cite[cf.][]{vanZee2004a}. Interestingly, already with the present
  compilation of measurements, a correlation with luminosity becomes
  evident (right panel in Fig.~\ref{rotfig}), with rotationally
  flattened galaxies being almost exclusively found among the brighter
  ($M_r\la-17$) objects \citep[also see][]{Ludwig2009}.


%


While obtaining dE rotation curves through integrated galaxy light
becomes extremely time-expensive at two or more half-light radii, the
globular cluster systems provide a worthwhile alternative. With their
velocities, galaxy dynamics can be mapped out to very large
radii. \citet{Beasley2009} have found significant rotation in three
dEs that show only little outer  rotation from their integrated
light. This could suggest that significant rotation might be present
in many more dEs than previously thought. The dynamical $B$-band
mass-to-light ($M/L$)
ratios derived by \citet{Beasley2009}, ranging from $\sim$5 to
$\sim$10, are comparable to those found for giant E/S0s. Similarly,
the dynamical $M/L$ values derived by \citet{Geha2003}, based on
kinematics in the inner galaxy regions, range from 3 to 6 in $V$, thus
being consistent with intermediate-age to old stellar
populations. Since these studies targeted bright dEs, their results
are consistent with the study of \citet[also this issue]{Penny2009}, who used a different
approach that does not rely on kinematical
measurements. \citeauthor{Penny2009} conclude that only the fainter
dEs require dark matter to remain stable within the cluster potential.

\section{Environmental relations}

\begin{figure}
\includegraphics[width=76mm]{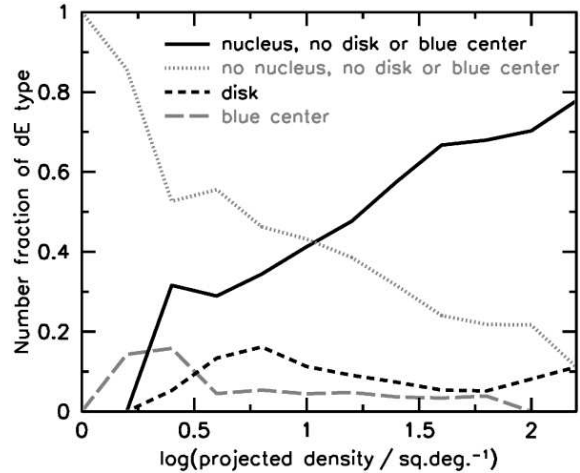}
\caption{Morphology-density relation \emph{within} the dE class: 
 number fractions of Virgo cluster dE subclasses (see
  Fig.~\ref{treefig}) with respect to local projected galaxy number density.}
\label{zoodensfig}
\end{figure}


 Among dwarf galaxies, the fraction of those with early-type
morphology is significantly larger in dynamically more evolved
environments \citep{Trentham2002}. Apart from this global trend,
early-type dwarfs also show a
pronounced relation with local galaxy density within a given
cluster or group: \citet{bin87} found that their number strongly
increases with local density, analogous to the correlation for
giants \citep{dre80}, and \citet{Tully2008} observed in (rich) galaxy groups
that early-type dwarfs strongly cluster around major E/S0 galaxies.

In addition, there is also a morphology-density relation \emph{within}
the dE class. As illustrated in Fig.~\ref{zoodensfig}, dEs with a blue
center and no nucleus, which have relatively flat shapes
(Fig.~\ref{treefig}), are preferentially found in the low-density 
cluster regions. Those with disks, which are the flattest subpopulation,
are located more towards intermediate-density regions, and nucleated
dEs (without disks or blue centers), which are the roundest dEs,
dominate the high-density regions. The different clustering properties
of nucleated and non-nucleated dEs
\citep[e.g.][]{vandenBergh1986,Ferguson1989} were challenged by
\citet{Cote2006}, who conjectured that this might only be due to a
selection bias, but \citet{p3} showed that no bias is present.

The stellar population properties also correlate with location in the
cluster. \citet{Smith2008} reported significantly younger ages of dEs
in the Coma cluster outskirts, and also higher metallicities. Similar
correlations are seen in Virgo, yet they appear not as
strong if dEs with disks are excluded \citep{Paudel2009}. These
findings agree with the UV/optical photometric study of Virgo dEs by
Kim et al.\ (this issue), who find bluer UV$-$optical colours
for dEs in the outskirts as compared to the center, and a
different, bluer CMR for dEs with disks than for those without disks.
This gives a consistent picture with the results of \citet{Toloba2009},
that rotationally flattened dEs are preferentially found in the
cluster outskirts, and are also younger than the non-rotating ones.

It is important to note that the scaling relations and the CMR of dEs
were found to be very similar in environments of different density,
from groups to clusters \citep{DeRijcke2009}. In this respect, it is
also noteworthy that\linebreak \citet{Tully2008} found a high
fraction of dEs with blue central star-forming regions in an
intermediate-mass group -- as compared to the rather small fraction in
the Virgo cluster -- and \citet{Gu2006} even presented an isolated dE
with such a blue core \citep[also see][]{Peeples2008}.

\section{Formation scenarios}

It has
become popular belief that dEs were formed at relatively late epochs by
environmental effects that structurally transformed spiral and
irregular galaxies \citep[e.g.][]{Moore1998}.
 At the same time, however, dEs are predicted 
to form in models of a $\Lambda$CDM universe, as the descendants of building
blocks in hierarchical structure formation \linebreak \citep[e.g.][]{Moore1999}.
Before discussing these scenarios in more detail, it might be useful to
summarize what a dE formation mechanism, or a
combined effect of several mechanisms, needs to produce: a galaxy with
(i) smoothly distributed starlight (Fig.~\ref{dEbigsmallfig}), (ii) a
SB profile whose slope is determined by the galaxy luminosity and is
not necessarily exponential (Fig.~\ref{profilefig}), (iii) a
nearly ellipsoidal shape, although allowing a rather large range of
axial ratios (Fig.~\ref{treefig}), (iv) little or no remaining gas
\citep{diSer07}, (v) a relatively old stellar population with at most only a
small contribution of young
stars (in terms of mass,\linebreak \citealt{p2}), nevertheless allowing a
large range in ages \citep{Paudel2009}. This list could of course be
continued with metallicity, kinematics, or further characteristics.


The simplest route to dEs would be if no modification of the shape
were necessary. This would merely require to remove the gas from the
galaxy (\emph{stripping}, e.g.\ through the ram-pressure of the hot
intracluster medium, \citealt{gun72}), as well as the cold gas
reservoir around the galaxy (\emph{starvation},
\citealt{Larson1980}). Both are typical processes of a cluster
environment and are not
restricted to the central region \citep{Tonnesen2007}.
At least for the non-nucleated dEs (dE(nN)s), star-forming galaxies
with similar shapes are the (dwarf) irregulars (Irrs; \citealt{Binggeli1995}). The median observed axial
ratios of bright and faint Virgo dE(nN)s \citep{p3} are
0.51 and 0.71, respectively, calculated from VCC values. When
subdividing the Irrs at their median $B$ magnitude, their
corresponding median axial ratios are 0.54 and 0.63. With a more
extreme definition of bright and faint, by excluding $\pm 0\fm50$
around the median magnitude, the values are 0.54 and 0.68. These
values, and the presence of a correlation with galaxy luminosity,
are very similar between Irrs and dE(nN)s, and might seem to imply
that dE(nN)s could indeed be stripped Irrs.

However, the direct evolution from Irrs to dEs faces some
problems: the Irrs have a too low metallicity \citep{thu85} and would
end up with a too low surface brightness after cessation of star
formation \citep{bot86,Boselli2008}. These problems can be overcome
by a scenario 
in which the initially lower metallicity and surface brightness of an
Irr are increased by several bursts of star formation \citep{dav88},
during which the galaxy appears as blue compact dwarf (BCD). After the
last BCD phase it fades to become a dE
\citep[cf.][]{Dellenbusch2008}. It can be regarded as support 
for this scenario that
\citet{Vaduvescu2006} and \citet{Vaduvescu2008} found dEs and BCDs to fall on the
``fundamental plane of dwarf irregulars'', a relatively tight relation
between near-infrared absolute $K$ magnitude, 
central surface brightness, and H\,I line width, substituted
by stellar velocity dispersion for the dEs.
The required bursts of star formation could be triggered by tidal
interactions \citep{Lavery1988,Moss1998}, leading to a rapid
conversion of gas into stars.

This indirect way of making a dE out of an Irr, by enhanced
consumption of gas, might seem more appealing than simple gas removal
\citep{Sabatini2005}: if those evolutionary paths leading to today's
Irrs would have been altered by ram-pressure stripping events that
removed the gas at some time in the past, the resulting dEs would be
even fainter (in total magnitude and surface brightness) than todays
Irrs \citep{Boselli2008}. Stripped Irrs -- or more precisely:
stripped galaxies that would otherwise have become today's Irrs --
could thus be the progenitors of fainter dEs or dSphs, but not of the
brighter ones. If the latter were to be explained by ram-pressure stripping,
the corresponding progenitors need to be spiral
(Sc/Sd) galaxies \citep{Boselli2008}. However, these galaxies, or at
least their disks, are much flatter than even the dEs with disks, so
pure stripping seems not sufficient. A plausible solution could be
that tidal interactions lead
to significant disk thickening \citep{Gnedin2003}. If they even lead
to the disruption of the disk, the bulge -- if present -- would
remain, and its properties might fall in the parameter space of dEs
\citep{Whitmore1993,Ferguson1994,Graham2008}. 

A similarly violent scenario is possible also with a bulgeless disk
entering a cluster:
repeated close encounters with massive galaxies can lead to
substantial mass loss and a complete structural transformation, resulting
in a dE \citep{Moore1996}. The simulations of
\citet{Mastropietro2005} showed that only those galaxies that end up on orbits
within the central cluster region experience such a strong
transformation. Those with more eccentric and/or outer orbits retain
more of their original disk structure, as well as a larger
$v/\sigma$. A further investigation of such fast galaxy-galaxy
interactions was performed by \citet{Aguerri2009}, who confirmed that
these processes are efficient mechanisms to transform late-type
disks galaxies into dEs, and whose simulated remnants are at least able
to populate part of the fundamental plane locus of dEs.
Since neither tidal interactions nor ram-pressure stripping are
avoidable for galaxies within a cluster, the real situation is
probably best described with an interplay of the different
 mechanisms.

At the same time, however, dEs are predicted 
to form in models of a $\Lambda$CDM universe, as descendants of building
blocks in hierarchical structure formation. Cosmological simulations
of galaxy clusters do predict a large number of dark matter subhaloes
that are appropriate to host dEs \citep[e.g.][]{Moore1999} ---
there is no ``missing satellites problem'' for clusters.
In that sense, dEs could
be close relatives to their giant counterparts, sharing a cosmological
origin. Indeed, semi-analytical models of galaxy formation,
which are tied to cosmological N-body simulations, are able to reproduce the
location of dEs on the fundamental plane when taking into account
the dynamical response after supernova-driven gas-loss \linebreak\citep{DeRijcke2005,Janz2008}. Since this
effect is mostly negligible at higher masses, it explains naturally
the apparently different behaviour of Es and dEs (see
Janz \& Lisker, this issue).


Further possibilities for dE formation exist. The merger of very
gas-rich dwarfs could lead to BCDs \citep{Bekki2008BCD}, which might
then become dEs as discussed earlier. Interactions of giant
spiral galaxies could form {\it tidal dwarfs} that become dEs, a
scenario that -- under certain assumptions -- could explain the number
of dEs in clusters\linebreak \citep{Okazaki2000}. All these scenarios could
ideally be investigated by comparing the measurements from
multiwavelength observations to models of the chemical and
structural evolution of dwarf galaxies or their progenitors
\citep[e.g.][]{Hensler2004,Boselli2008,Valcke2008}.



\section{Discussion}

\subsection{Early versus late formation}

Most dEs in cluster regions of high density are pressure supported
(Fig.~\ref{rotfig}), whereas in intermediate and low-density regions, more
galaxies seem to be flattened by rotation or be rotationally
supported\linebreak \citep{vanZee2004a,Toloba2009}. Does this, by itself, tell us something
about the formation
history of those dEs? Simulations show that relatively flat
low-mass cluster galaxies experiencing repeated strong tidal
interactions become dynamically hotter: their $v/\sigma$
decreases and they become ``rounder'', i.e.\ their axial ratio increases\linebreak
\citep{Mastropietro2005}. Such interactions are 
more frequent in the central region of a cluster. Thus, if all dEs
once had similar levels of rotational support, those in the central
cluster region would have become dynamically hotter, in agreement with
the observational findings (I call this {\it case 1}). Likewise, if
those that are now closer to the center have resided in the 
cluster for a longer time, they experienced more such interactions,
leading to the same result ({\it case 2}).
Similar scenarios apply to the observation that galaxies in the
central cluster regions have higher stellar population ages
on average \citep{Smith2008}: ram-pressure stripping, and also tidal stripping,
are stronger in the central region, thus quenching star formation
earlier in the galaxies that reside there. But without
further considerations, it cannot be decided whether those galaxies
have spent a longer time in the cluster, and were therefore stripped
earlier (case 2), or
whether they were simply stripped more efficiently (case 1).

From case 2, one might be tempted to conclude a later infall and
transformation of the progenitors of those dEs that are now in the
cluster outskirts and are significantly flattened by
rotation. However, case 1 does not imply anything about when
the dEs, or their progenitors, 
have entered the cluster, or whether they have always been part of the
cluster since its formation phase. This is the point at which input
from the theoretical side is essential: for case 2, we need
to have an understanding of the timescale between the infall of a
group of progenitor galaxies and the point when these galaxies (or
their remnants) reach a stage in which they are concentrated towards
the cluster center and show a peaked and fairly regular distribution of
heliocentric velocities. For case 1, probably even more complex input physics is needed,
since semi-analytical model predictions for ``cosmological dwarfs''
that form in dense environments need to be tested against the
observations. Equally important are observational studies of dEs at
significantly earlier epochs, and comparisons to the local population
(cf.\ \citealt{Andreon2008}; \citealt{Harsono2009}, also this issue).

\subsection{Multiple origins?}

Would it be possible, despite the complex characteristics and several
different subclasses, to explain all dEs by just a single formation
channel? One could, for example, imagine that dEs with disks (dE(di)s)
do not form an intrinsically different subclass, but instead they
simply constitute the flat tail of the other dEs, with their disk components
having not been destroyed yet. The fraction of dE(di)s with nuclei
($\sim$75\%, \citealt{p4}) agrees with  the overall fraction
of nucleated dEs in the (bright) magnitude range of the dE(di)s.
However, while bright non-nucleated dEs (without disks or
blue centers) are significantly flatter than faint ones, this is not
the case for the nucleated dEs (again without disks or blue
centers). Given the lower stellar population age of dE(nN)s
\citep{Rakos2004}, and their 
location in regions of lower density, a possible
explanation would be that the brighter dE(nN)s still need to
experience further dynamical heating through tidal interactions,
before they become as round as the dE(N)s. For the faint dE(nN)s, though,
one would have to assume that they were more susceptible to such
interactions, and therefore already have a rounder shape. On the other
hand, the correlation of the projected shape of dE(N)s and their
heliocentric velocity \citep{Lisker2009velo} includes both bright and
faint objects, and seems to imply that the roundest shapes are
achieved only when the galaxies' orbits are significantly circularized.

To pursue the ``unification'' of dE subclasses, another
requirement is that the dE(nN)s would need to form nuclei rather soon,
before their stellar population ages are comparable to today's
dE(N)s. (Or, alternatively, there might have been some mechanism in
the past that supported nucleus formation, but is not as
efficient anymore today.) Perhaps nuclei are currently being formed in
the centers of the dEs with blue cores, where we are
witnessing ongoing star formation \citep{p2}. This would lead to
nuclei with younger stellar populations than those of their host
galaxies. However, \citet{Cote2006} found dE nuclei to have
intermediate to old ages, and \citet{Lotz2004} measured similar
colours of nuclei and dE globular clusters (GCs). Alternatively, most
nuclei could form through coalescence of GCs, which would indeed be a
more efficient process in the central cluster regions
\citep{Oh2000}. Still, some effect would have to explain why the ratio
of dE(N)s and dE(nN)s increases strongly with increasing
dE luminosity \citep{san85b}, especially if the dE(nN)s were to be the
immediate progenitors of the dE(N)s.

Similar to the above argument about nucleus formation, the spatial
distribution of dE(nN)s would soon have to become significantly more
centrally concentrated.\linebreak \citet{Conselice2001} derived a two-body
relaxation time for Virgo dEs of much more than a Hubble time.
However, two-body relaxation might be too simple for
the real situation. These issues are probably best
investigated with cosmological simulations, by ``flagging'' different
groups or individual galaxies when they enter a cluster, and following
the evolution of their combined (observable) state over various
timescales. Current and future generations of simulations\linebreak
\citep[e.g.][]{BoylanKolchin2009} certainly provide such a
possibility.

Obviously, many ifs and thens are necessary to explain all dEs
by a single formation channel, and several authors have argued in
favour of multiple channels
(\citealt{p4}; \citealt{Poggianti2001}; \citealt{Rakos2004};\linebreak \citealt{vanZee2004b}). Nevertheless, several
aspects of dEs still need to be explored both theoretically and
observationally, like the presence and characteristics of their
globular clusters and what they can tell us about their evolutionary
history \citep[see the discussion in][]{Boselli2008}.

\subsection{A plea for the study of early-type dwarfs}

Galaxy mass is one of the main drivers of galaxy evolution and
appearance. Internal physical processes like supernova feedback \citep{DekelSilk1986}
or the dynamical response to gas loss \citep{YoshiiArimoto1987}
 have a stronger impact on dwarf galaxies
than on giants, making dwarfs valuable test objects for galaxy
formation models\linebreak \citep{Janz2008,Janz2009a}. Nevertheless, if we want
to use them as probes of the physical mechanisms that govern galaxy
evolution, we need to understand the origin(s) of this  most abundant
galaxy population of clusters --- which is a difficult task.
The intriguing complexity of dEs is
contrasted with the moderate amount of high-quality observational data.
Now that the largest cosmological simulations and
their semi-analytic models begin to reach significantly into
the dwarf regime, it is essential to build complete
observational samples providing a thorough characterization of fundamental
scaling relations at low masses, involving structure, kinematics, and
stellar population properties. These will provide indispensable
benchmarks for new generations of galaxy models, as well as for future
studies of dEs at significantly earlier epochs with extremely large
telescopes.

\acknowledgements

I would like to thank the organizers for a successful symposium, Jan
Pflamm-Altenburg and Pavel Kroupa for providing Fig.~\ref{dabfig}, 
and Heikki Salo for helpful discussions.
I am supported within the framework of the Excellence Initiative by
the German Research Foundation (DFG) through the Heidelberg Graduate
School of Fundamental Physics (grant number GSC 129/1). 
The images for Figs.~\ref{dEbigsmallfig} and \ref{dEspiral} were
collected at the European Organisation for Astronomical Research in the Southern Hemisphere,
    Chile, for programme 077.B-0785 (P.I.\ T.\ Lisker).



\begin{thebibliography}{126}

\bibitem[{{Adami} {et~al.}(2006){Adami}, {Scheidegger}, {Ulmer}, {Durret},
  {Mazure}, {West}, {Conselice}, {Gregg}, {Kasun}, {Pell{\'o}}, \&
  {Picat}}]{Adami2006}
{Adami}, C., {Scheidegger}, R., {Ulmer}, M., {et~al.} 2006, \aap, 459, 679

\bibitem[{{Aguerri} \& {Gonz{\'a}lez-Garc{\'{\i}}a}(2009)}]{Aguerri2009}
{Aguerri}, J.~A.~L. \& {Gonz{\'a}lez-Garc{\'{\i}}a}, A.~C. 2009, \aap, 494, 891

\bibitem[{{Aguerri} {et~al.}(2005){Aguerri}, {Iglesias-P{\' a}ramo},
  {V{\'{\i}}lchez}, {Mu{\~ n}oz-Tu{\~ n}{\' o}n}, \& {S{\'
  a}nchez-Janssen}}]{Aguerri2005}
{Aguerri}, J.~A.~L., {Iglesias-P{\' a}ramo}, J., {V{\'{\i}}lchez}, J.~M.,
  {Mu{\~ n}oz-Tu{\~ n}{\' o}n}, C., \& {S{\' a}nchez-Janssen}, R. 2005, \aj,
  130, 475

\bibitem[{{Andreon}(2008)}]{Andreon2008}
{Andreon}, S. 2008, \mnras, 386, 1045

\bibitem[{{Andreon} {et~al.}(2006){Andreon}, {Cuillandre}, {Puddu}, \&
  {Mellier}}]{Andreon2006}
{Andreon}, S., {Cuillandre}, J.-C., {Puddu}, E., \& {Mellier}, Y. 2006, \mnras,
  372, 60

\bibitem[{{Barazza} {et~al.}(2002){Barazza}, {Binggeli}, \&
  {Jerjen}}]{Barazza2002a}
{Barazza}, F.~D., {Binggeli}, B., \& {Jerjen}, H. 2002, \aap, 391, 823

\bibitem[{{Beasley} {et~al.}(2009){Beasley}, {Cenarro}, {Strader}, \&
  {Brodie}}]{Beasley2009}
{Beasley}, M.~A., {Cenarro}, A.~J., {Strader}, J., \& {Brodie}, J.~P. 2009,
  \aj, 137, 5146

\bibitem[{{Bekki}(2008)}]{Bekki2008BCD}
{Bekki}, K. 2008, \mnras, 388, L10

\bibitem[{{Bender} \& {Nieto}(1990)}]{BenderNieto1990}
{Bender}, R. \& {Nieto}, J.-L. 1990, \aap, 239, 97

\bibitem[{{Binggeli} \& {Cameron}(1991)}]{bin91}
{Binggeli}, B. \& {Cameron}, L.~M. 1991, \aap, 252, 27

\bibitem[{{Binggeli} \& {Jerjen}(1998)}]{Binggeli1998}
{Binggeli}, B. \& {Jerjen}, H. 1998, \aap, 333, 17

\bibitem[{{Binggeli} \& {Popescu}(1995)}]{Binggeli1995}
{Binggeli}, B. \& {Popescu}, C.~C. 1995, \aap, 298, 63

\bibitem[{{Binggeli} {et~al.}(1985){Binggeli}, {Sandage}, \& {Tammann}}]{vcc}
{Binggeli}, B., {Sandage}, A., \& {Tammann}, G.~A. 1985, \aj, 90, 1681

\bibitem[{{Binggeli} {et~al.}(1987){Binggeli}, {Tammann}, \& {Sandage}}]{bin87}
{Binggeli}, B., {Tammann}, G.~A., \& {Sandage}, A. 1987, \aj, 94, 251

\bibitem[{{Binney}(1978)}]{Binney1978}
{Binney}, J. 1978, \mnras, 183, 501

\bibitem[{{Boselli} {et~al.}(2008){Boselli}, {Boissier}, {Cortese}, \&
  {Gavazzi}}]{Boselli2008}
{Boselli}, A., {Boissier}, S., {Cortese}, L., \& {Gavazzi}, G. 2008, \apj, 674,
  742

\bibitem[{{Boselli} {et~al.}(2005){Boselli}, {Cortese}, {Deharveng}, {Gavazzi},
  {Yi}, {Gil de Paz}, {Seibert}, {Boissier}, {Donas}, {Lee}, {Madore},
  {Martin}, {Rich}, \& {Sohn}}]{Boselli2005}
{Boselli}, A., {Cortese}, L., {Deharveng}, J.~M., {et~al.} 2005, \apjl, 629,
  L29

\bibitem[{{Bothun} {et~al.}(1986){Bothun}, {Mould}, {Caldwell}, \&
  {MacGillivray}}]{bot86}
{Bothun}, G.~D., {Mould}, J.~R., {Caldwell}, N., \& {MacGillivray}, H.~T. 1986,
  \aj, 92, 1007

\bibitem[{{Boylan-Kolchin} {et~al.}(2009){Boylan-Kolchin}, {Springel}, {White},
  {Jenkins}, \& {Lemson}}]{BoylanKolchin2009}
{Boylan-Kolchin}, M., {Springel}, V., {White}, S.~D.~M., {Jenkins}, A., \&
  {Lemson}, G. 2009, \mnras, 398, 1150

\bibitem[{{Caldwell}(1983)}]{cal83}
{Caldwell}, N. 1983, \aj, 88, 804

\bibitem[{{Caldwell}(2006)}]{Caldwell2006}
{Caldwell}, N. 2006, \apj, 651, 822

\bibitem[{{Chang} {et~al.}(2006){Chang}, {Gallazzi}, {Kauffmann}, {Charlot},
  {Ivezi{\'c}}, {Brinchmann}, \& {Heckman}}]{cha06}
{Chang}, R., {Gallazzi}, A., {Kauffmann}, G., {et~al.} 2006, \mnras, 366, 717

\bibitem[{{Chilingarian}(2009)}]{Chilingarian2009}
{Chilingarian}, I.~V. 2009, \mnras, 394, 1229

\bibitem[{{Conselice} {et~al.}(2001){Conselice}, {Gallagher}, \&
  {Wyse}}]{Conselice2001}
{Conselice}, C.~J., {Gallagher}, III, J.~S., \& {Wyse}, R.~F.~G. 2001, \apj,
  559, 791

\bibitem[{{Conselice} {et~al.}(2003){Conselice}, {Gallagher}, \&
  {Wyse}}]{conIII}
{Conselice}, C.~J., {Gallagher}, III, J.~S., \& {Wyse}, R.~F.~G. 2003, \aj,
  125, 66

\bibitem[{{C{\^o}t{\'e}} {et~al.}(2007){C{\^o}t{\'e}}, {Ferrarese},
  {Jord{\'a}n}, {Blakeslee}, {Chen}, {Infante}, {Merritt}, {Mei}, {Peng},
  {Tonry}, {West}, \& {West}}]{Cote2007}
{C{\^o}t{\'e}}, P., {Ferrarese}, L., {Jord{\'a}n}, A., {et~al.} 2007, \apj,
  671, 1456

\bibitem[{{C{\^o}t{\'e}} {et~al.}(2006){C{\^o}t{\'e}}, {Piatek}, {Ferrarese},
  {Jord{\'a}n}, {Merritt}, {Peng}, {Ha{\c s}egan}, {Blakeslee}, {Mei}, {West},
  {Milosavljevi{\'c}}, \& {Tonry}}]{Cote2006}
{C{\^o}t{\'e}}, P., {Piatek}, S., {Ferrarese}, L., {et~al.} 2006, \apjs, 165,
  57

\bibitem[{{Dabringhausen} {et~al.}(2008){Dabringhausen}, {Hilker}, \&
  {Kroupa}}]{Dabringhausen2008}
{Dabringhausen}, J., {Hilker}, M., \& {Kroupa}, P. 2008, \mnras, 386, 864

\bibitem[{{Davies} \& {Phillipps}(1988)}]{dav88}
{Davies}, J.~I. \& {Phillipps}, S. 1988, \mnras, 233, 553

\bibitem[{{De Rijcke} {et~al.}(2005){De Rijcke}, {Michielsen}, {Dejonghe},
  {Zeilinger}, \& {Hau}}]{DeRijcke2005}
{De Rijcke}, S., {Michielsen}, D., {Dejonghe}, H., {Zeilinger}, W.~W., \&
  {Hau}, G.~K.~T. 2005, \aap, 438, 491

\bibitem[{{De Rijcke} {et~al.}(2009){De Rijcke}, {Penny}, {Conselice},
  {Valcke}, \& {Held}}]{DeRijcke2009}
{De Rijcke}, S., {Penny}, S.~J., {Conselice}, C.~J., {Valcke}, S., \& {Held},
  E.~V. 2009, \mnras, 393, 798

\bibitem[{{De Rijcke} {et~al.}(2007){De Rijcke}, {Zeilinger}, {Hau},
  {Prugniel}, \& {Dejonghe}}]{DeRijcke2007}
{De Rijcke}, S., {Zeilinger}, W.~W., {Hau}, G.~K.~T., {Prugniel}, P., \&
  {Dejonghe}, H. 2007, \apj, 659, 1172

\bibitem[{{de Vaucouleurs}(1961)}]{deV61}
{de Vaucouleurs}, G. 1961, \apjs, 5, 233

\bibitem[{{Dekel} \& {Silk}(1986)}]{DekelSilk1986}
{Dekel}, A. \& {Silk}, J. 1986, \apj, 303, 39

\bibitem[{{Dellenbusch} {et~al.}(2008){Dellenbusch}, {Gallagher}, {Knezek}, \&
  {Noble}}]{Dellenbusch2008}
{Dellenbusch}, K.~E., {Gallagher}, III, J.~S., {Knezek}, P.~M., \& {Noble},
  A.~G. 2008, \aj, 135, 326

\bibitem[{{di Serego Alighieri} {et~al.}(2007){di Serego Alighieri}, {Gavazzi},
  {Giovanardi}, {Giovanelli}, {Grossi}, {Haynes}, {Kent}, {Koopmann},
  {Pellegrini}, {Scodeggio}, \& {Trinchieri}}]{diSer07}
{di Serego Alighieri}, S., {Gavazzi}, G., {Giovanardi}, C., {et~al.} 2007,
  \aap, 474, 851

\bibitem[{{Dressler}(1980)}]{dre80}
{Dressler}, A. 1980, \apj, 236, 351

\bibitem[{{Durrell} {et~al.}(2007){Durrell}, {Williams}, {Ciardullo},
  {Feldmeier}, {von Hippel}, {Sigurdsson}, {Jacoby}, {Ferguson}, {Tanvir},
  {Arnaboldi}, {Gerhard}, {Aguerri}, {Freeman}, \& {Vinciguerra}}]{Durrell2007}
{Durrell}, P.~R., {Williams}, B.~F., {Ciardullo}, R., {et~al.} 2007, \apj, 656,
  746

\bibitem[{{Faber}(1973)}]{fab73}
{Faber}, S.~M. 1973, \apj, 179, 731

\bibitem[{{Faber} \& {Jackson}(1976)}]{FaberJackson}
{Faber}, S.~M. \& {Jackson}, R.~E. 1976, \apj, 204, 668

\bibitem[{{Ferguson} \& {Binggeli}(1994)}]{Ferguson1994}
{Ferguson}, H.~C. \& {Binggeli}, B. 1994, \aapr, 6, 67

\bibitem[{{Ferguson} \& {Sandage}(1989)}]{Ferguson1989}
{Ferguson}, H.~C. \& {Sandage}, A. 1989, \apjl, 346, L53

\bibitem[{{Ferrarese} {et~al.}(2006){Ferrarese}, {C{\^o}t{\'e}}, {Jord{\'a}n},
  {Peng}, {Blakeslee}, {Piatek}, {Mei}, {Merritt}, {Milosavljevi{\'c}},
  {Tonry}, \& {West}}]{Ferrarese06}
{Ferrarese}, L., {C{\^o}t{\'e}}, P., {Jord{\'a}n}, A., {et~al.} 2006, \apjs,
  164, 334

\bibitem[{{Gavazzi} {et~al.}(2002){Gavazzi}, {Bonfanti}, {Sanvito}, {Boselli},
  \& {Scodeggio}}]{gavazzi2002}
{Gavazzi}, G., {Bonfanti}, C., {Sanvito}, G., {Boselli}, A., \& {Scodeggio}, M.
  2002, \apj, 576, 135

\bibitem[{{Gavazzi} {et~al.}(2005){Gavazzi}, {Donati}, {Cucciati}, {Sabatini},
  {Boselli}, {Davies}, \& {Zibetti}}]{Gavazzi2005a}
{Gavazzi}, G., {Donati}, A., {Cucciati}, O., {et~al.} 2005, \aap, 430, 411

\bibitem[{{Geha} {et~al.}(2002){Geha}, {Guhathakurta}, \& {van der
  Marel}}]{Geha2002}
{Geha}, M., {Guhathakurta}, P., \& {van der Marel}, R.~P. 2002, \aj, 124, 3073

\bibitem[{{Geha} {et~al.}(2003){Geha}, {Guhathakurta}, \& {van der
  Marel}}]{Geha2003}
{Geha}, M., {Guhathakurta}, P., \& {van der Marel}, R.~P. 2003, \aj, 126, 1794

\bibitem[{{Gil de Paz} {et~al.}(2007){Gil de Paz}, {Boissier}, {Madore},
  {Seibert}, {Joe}, {Boselli}, {Wyder}, {Thilker}, {Bianchi}, {Rey}, {Rich},
  {Barlow}, {Conrow}, {Forster}, {Friedman}, {Martin}, {Morrissey}, {Neff},
  {Schiminovich}, {Small}, {Donas}, {Heckman}, {Lee}, {Milliard}, {Szalay}, \&
  {Yi}}]{GildePaz2007}
{Gil de Paz}, A., {Boissier}, S., {Madore}, B.~F., {et~al.} 2007, \apjs, 173,
  185

\bibitem[{{Gnedin}(2003)}]{Gnedin2003}
{Gnedin}, O.~Y. 2003, \apj, 589, 752

\bibitem[{{Graham}(2002)}]{GrahamM32}
{Graham}, A.~W. 2002, \apjl, 568, L13

\bibitem[{{Graham} \& {Guzm{\'a}n}(2003)}]{Graham2003a}
{Graham}, A.~W. \& {Guzm{\'a}n}, R. 2003, \aj, 125, 2936

\bibitem[{{Graham} {et~al.}(2003){Graham}, {Jerjen}, \& {Guzm{\'
  a}n}}]{Graham2003b}
{Graham}, A.~W., {Jerjen}, H., \& {Guzm{\' a}n}, R. 2003, \aj, 126, 1787

\bibitem[{{Graham} \& {Worley}(2008)}]{Graham2008}
{Graham}, A.~W. \& {Worley}, C.~C. 2008, \mnras, 388, 1708

\bibitem[{{Grant} {et~al.}(2005){Grant}, {Kuipers}, \& {Phillipps}}]{Grant2005}
{Grant}, N.~I., {Kuipers}, J.~A., \& {Phillipps}, S. 2005, \mnras, 363, 1019

\bibitem[{{Gu} {et~al.}(2006){Gu}, {Zhao}, {Shi}, {Peng}, \& {Luo}}]{Gu2006}
{Gu}, Q., {Zhao}, Y., {Shi}, L., {Peng}, Z., \& {Luo}, X. 2006, \aj, 131, 806

\bibitem[{{Gunn} \& {Gott}(1972)}]{gun72}
{Gunn}, J.~E. \& {Gott}, J.~R.~I. 1972, \apj, 176, 1

\bibitem[{{Harsono} \& {DePropris}(2009)}]{Harsono2009}
{Harsono}, D. \& {DePropris}, R. 2009, \aj, 137, 3091

\bibitem[{{Held} \& {Mould}(1994)}]{held1994}
{Held}, E.~V. \& {Mould}, J.~R. 1994, \aj, 107, 1307

\bibitem[{{Hensler} {et~al.}(2004){Hensler}, {Theis}, \&
  {Gallagher}}]{Hensler2004}
{Hensler}, G., {Theis}, C., \& {Gallagher}, J.~S., I. 2004, \aap, 426, 25

\bibitem[{{Janz} \& {Lisker}(2008)}]{Janz2008}
{Janz}, J. \& {Lisker}, T. 2008, \apjl, 689, L25

\bibitem[{{Janz} \& {Lisker}(2009)}]{Janz2009a}
{Janz}, J. \& {Lisker}, T. 2009, \apjl, 696, L102

\bibitem[{{Jerjen} {et~al.}(2000){Jerjen}, {Kalnajs}, \&
  {Binggeli}}]{Jerjen2000a}
{Jerjen}, H., {Kalnajs}, A., \& {Binggeli}, B. 2000, \aap, 358, 845

\bibitem[{{Jerjen} \& {Tammann}(1997)}]{Jerjen1997}
{Jerjen}, H. \& {Tammann}, G.~A. 1997, \aap, 321, 713

\bibitem[{{Kodama} \& {Arimoto}(1997)}]{kod97}
{Kodama}, T. \& {Arimoto}, N. 1997, \aap, 320, 41

\bibitem[{{Koleva} {et~al.}(2009){Koleva}, {De Rijcke}, {Prugniel},
  {Zeilinger}, \& {Michielsen}}]{Koleva2009}
{Koleva}, M., {De Rijcke}, S., {Prugniel}, P., {Zeilinger}, W.~W., \&
  {Michielsen}, D. 2009, \mnras, 396, 2133

\bibitem[{{Kormendy}(1982)}]{Kormendy1982proc}
{Kormendy}, J. 1982, in Morphology and Dynamics of
  Galaxies, ed. {L.~Martinet \& M.~Mayor}, 113

\bibitem[{{Kormendy} {et~al.}(2009){Kormendy}, {Fisher}, {Cornell}, \&
  {Bender}}]{Kormendy2009}
{Kormendy}, J., {Fisher}, D.~B., {Cornell}, M.~E., \& {Bender}, R. 2009, \apjs,
  182, 216

\bibitem[{{Larson} {et~al.}(1980){Larson}, {Tinsley}, \&
  {Caldwell}}]{Larson1980}
{Larson}, R.~B., {Tinsley}, B.~M., \& {Caldwell}, C.~N. 1980, \apj, 237, 692

\bibitem[{{Lavery} \& {Henry}(1988)}]{Lavery1988}
{Lavery}, R.~J. \& {Henry}, J.~P. 1988, \apj, 330, 596

\bibitem[{{Lisker} \& {Fuchs}(2009)}]{LiskerFuchs2009}
{Lisker}, T. \& {Fuchs}, B. 2009, \aap, 501, 429

\bibitem[{{Lisker} {et~al.}(2006{\natexlab{b}}){Lisker}, {Glatt}, {Westera}, \&
  {Grebel}}]{p2}
{Lisker}, T., {Glatt}, K., {Westera}, P., \& {Grebel}, E.~K.
  2006{\natexlab{b}}, \aj, 132, 2432

\bibitem[{{Lisker} {et~al.}(2006{\natexlab{a}}){Lisker}, {Grebel}, \&
  {Binggeli}}]{p1}
{Lisker}, T., {Grebel}, E.~K., \& {Binggeli}, B. 2006{\natexlab{a}}, \aj, 132,
  497

\bibitem[{{Lisker} {et~al.}(2008){Lisker}, {Grebel}, \& {Binggeli}}]{p4}
{Lisker}, T., {Grebel}, E.~K., \& {Binggeli}, B. 2008, \aj, 135, 380

\bibitem[{{Lisker} {et~al.}(2007){Lisker}, {Grebel}, {Binggeli}, \&
  {Glatt}}]{p3}
{Lisker}, T., {Grebel}, E.~K., {Binggeli}, B., \& {Glatt}, K. 2007, \apj, 660,
  1186

\bibitem[{{Lisker} \& {Han}(2008)}]{LiskerHan2008}
{Lisker}, T. \& {Han}, Z. 2008, \apj, 680, 1042

\bibitem[{{Lisker} {et~al.}(2009){Lisker}, {Janz}, {Hensler}, {Kim}, {Rey},
  {Weinmann}, {Mastropietro}, {Hielscher}, {Paudel}, \&
  {Kotulla}}]{Lisker2009velo}
{Lisker}, T., {Janz}, J., {Hensler}, G., {et~al.} 2009, \apjl, submitted

\bibitem[{{Lotz} {et~al.}(2004){Lotz}, {Miller}, \& {Ferguson}}]{Lotz2004}
{Lotz}, J.~M., {Miller}, B.~W., \& {Ferguson}, H.~C. 2004, \apj, 613, 262

\bibitem[{Ludwig(2009)}]{Ludwig2009}
Ludwig, J. 2009, Diploma thesis, University of Heidelberg

\bibitem[{{Martin} {et~al.}(2005){Martin}, {Fanson}, {Schiminovich},
  {Morrissey}, {Friedman}, {Barlow}, {Conrow}, {Grange}, {Jelinsky},
  {Milliard}, {Siegmund}, {Bianchi}, {Byun}, {Donas}, {Forster}, {Heckman},
  {Lee}, {Madore}, {Malina}, {Neff}, {Rich}, {Small}, {Surber}, {Szalay},
  {Welsh}, \& {Wyder}}]{GALEX}
{Martin}, D.~C., {Fanson}, J., {Schiminovich}, D., {et~al.} 2005, \apjl, 619,
  L1

\bibitem[{{Mastropietro} {et~al.}(2005){Mastropietro}, {Moore}, {Mayer},
  {Debattista}, {Piffaretti}, \& {Stadel}}]{Mastropietro2005}
{Mastropietro}, C., {Moore}, B., {Mayer}, L., {et~al.} 2005, \mnras, 364, 607

\bibitem[{{Matkovi{\'c}} \& {Guzm{\'a}n}(2005)}]{mat05}
{Matkovi{\'c}}, A. \& {Guzm{\'a}n}, R. 2005, \mnras, 362, 289

\bibitem[{{Michielsen} {et~al.}(2008){Michielsen}, {Boselli}, {Conselice},
  {Toloba}, {Whiley}, {Arag{\'o}n-Salamanca}, {Balcells}, {Cardiel}, {Cenarro},
  {Gorgas}, {Peletier}, \& {Vazdekis}}]{Michielsen2008}
{Michielsen}, D., {Boselli}, A., {Conselice}, C.~J., {et~al.} 2008, \mnras,
  385, 1374

\bibitem[{{Misgeld} {et~al.}(2009){Misgeld}, {Hilker}, \&
  {Mieske}}]{Misgeld2009}
{Misgeld}, I., {Hilker}, M., \& {Mieske}, S. 2009, \aap, 496, 683

\bibitem[{{Misgeld} {et~al.}(2008){Misgeld}, {Mieske}, \&
  {Hilker}}]{Misgeld2008}
{Misgeld}, I., {Mieske}, S., \& {Hilker}, M. 2008, \aap, 486, 697

\bibitem[{{Moore} {et~al.}(1999){Moore}, {Ghigna}, {Governato}, {Lake},
  {Quinn}, {Stadel}, \& {Tozzi}}]{Moore1999}
{Moore}, B., {Ghigna}, S., {Governato}, F., {et~al.} 1999, \apjl, 524, L19

\bibitem[{{Moore} {et~al.}(1996){Moore}, {Katz}, {Lake}, {Dressler}, \&
  {Oemler}}]{Moore1996}
{Moore}, B., {Katz}, N., {Lake}, G., {Dressler}, A., \& {Oemler}, A. 1996,
  \nat, 379, 613

\bibitem[{{Moore} {et~al.}(1998){Moore}, {Lake}, \& {Katz}}]{Moore1998}
{Moore}, B., {Lake}, G., \& {Katz}, N. 1998, \apj, 495, 139

\bibitem[{{Moss} {et~al.}(1998){Moss}, {Whittle}, \& {Pesce}}]{Moss1998}
{Moss}, C., {Whittle}, M., \& {Pesce}, J.~E. 1998, \mnras, 300, 205

\bibitem[{{Oh} \& {Lin}(2000)}]{Oh2000}
{Oh}, K.~S. \& {Lin}, D.~N.~C. 2000, \apj, 543, 620

\bibitem[{{Okazaki} \& {Taniguchi}(2000)}]{Okazaki2000}
{Okazaki}, T. \& {Taniguchi}, Y. 2000, \apj, 543, 149

\bibitem[{{Paudel} {et~al.}(2009){Paudel}, {Lisker}, {Kuntschner}, {Grebel}, \&
  {Glatt}}]{Paudel2009}
{Paudel}, S., {Lisker}, T., {Kuntschner}, H., {Grebel}, E.~K., \& {Glatt}, K.
  2009, \mnras, submitted

\bibitem[{{Pedraz} {et~al.}(2002){Pedraz}, {Gorgas}, {Cardiel},
  {S{\'a}nchez-Bl{\'a}zquez}, \& {Guzm{\'a}n}}]{Pedraz2002}
{Pedraz}, S., {Gorgas}, J., {Cardiel}, N., {S{\'a}nchez-Bl{\'a}zquez}, P., \&
  {Guzm{\'a}n}, R. 2002, \mnras, 332, L59

\bibitem[{{Peeples} {et~al.}(2008){Peeples}, {Pogge}, \&
  {Stanek}}]{Peeples2008}
{Peeples}, M.~S., {Pogge}, R.~W., \& {Stanek}, K.~Z. 2008, \apj, 685, 904

\bibitem[{{Peng} {et~al.}(2006){Peng}, {Jord{\'a}n}, {C{\^o}t{\'e}},
  {Blakeslee}, {Ferrarese}, {Mei}, {West}, {Merritt}, {Milosavljevi{\'c}}, \&
  {Tonry}}]{Peng2006}
{Peng}, E.~W., {Jord{\'a}n}, A., {C{\^o}t{\'e}}, P., {et~al.} 2006, \apj, 639,
  95

\bibitem[{{Peng} {et~al.}(2008){Peng}, {Jord{\'a}n}, {C{\^o}t{\'e}},
  {Takamiya}, {West}, {Blakeslee}, {Chen}, {Ferrarese}, {Mei}, {Tonry}, \&
  {West}}]{Peng2008}
{Peng}, E.~W., {Jord{\'a}n}, A., {C{\^o}t{\'e}}, P., {et~al.} 2008, \apj, 681,
  197

\bibitem[{{Penny} \& {Conselice}(2008)}]{Penny2008}
{Penny}, S.~J. \& {Conselice}, C.~J. 2008, \mnras, 383, 247

\bibitem[{{Penny} {et~al.}(2009){Penny}, {Conselice}, {De Rijcke}, \&
  {Held}}]{Penny2009}
{Penny}, S.~J., {Conselice}, C.~J., {De Rijcke}, S., \& {Held}, E.~V. 2009,
  \mnras, 393, 1054

\bibitem[{{Pflamm-Altenburg} \& {Kroupa}(2009)}]{Pflamm2009}
{Pflamm-Altenburg}, J. \& {Kroupa}, P. 2009, \mnras, 397, 488

\bibitem[{{Poggianti} {et~al.}(2001){Poggianti}, {Bridges}, {Mobasher},
  {Carter}, {Doi}, {Iye}, {Kashikawa}, {Komiyama}, {Okamura}, {Sekiguchi},
  {Shimasaku}, {Yagi}, \& {Yasuda}}]{Poggianti2001}
{Poggianti}, B.~M., {Bridges}, T.~J., {Mobasher}, B., {et~al.} 2001, \apj, 562,
  689

\bibitem[{{Prugniel} \& {Simien}(1996)}]{Prugniel1996}
{Prugniel}, P. \& {Simien}, F. 1996, \aap, 309, 749

\bibitem[{{Rakos} \& {Schombert}(2004)}]{Rakos2004}
{Rakos}, K. \& {Schombert}, J. 2004, \aj, 127, 1502

\bibitem[{{Ryden} \& {Terndrup}(1994)}]{Ryden1994}
{Ryden}, B.~S. \& {Terndrup}, D.~M. 1994, \apj, 425, 43

\bibitem[{{Sabatini} {et~al.}(2003){Sabatini}, {Davies}, {Scaramella}, {Smith},
  {Baes}, {Linder}, {Roberts}, \& {Testa}}]{Sabatini2003}
{Sabatini}, S., {Davies}, J., {Scaramella}, R., {et~al.} 2003, \mnras, 341, 981

\bibitem[{{Sabatini} {et~al.}(2005){Sabatini}, {Davies}, {van Driel}, {Baes},
  {Roberts}, {Smith}, {Linder}, \& {O'Neil}}]{Sabatini2005}
{Sabatini}, S., {Davies}, J., {van Driel}, W., {et~al.} 2005, \mnras, 357, 819

\bibitem[{{Sandage} \& {Binggeli}(1984)}]{san84}
{Sandage}, A. \& {Binggeli}, B. 1984, \aj, 89, 919

\bibitem[{{Sandage} {et~al.}(1985){Sandage}, {Binggeli}, \& {Tammann}}]{san85b}
{Sandage}, A., {Binggeli}, B., \& {Tammann}, G.~A. 1985, \aj, 90, 1759

\bibitem[{{Sandage} \& {Visvanathan}(1978)}]{sanvis78a}
{Sandage}, A. \& {Visvanathan}, N. 1978, \apj, 223, 707

\bibitem[{{S{\'e}rsic}(1963)}]{Sersic1963}
{S{\'e}rsic}, J.~L. 1963, Boletin de la Asociacion Argentina de Astronomia La
  Plata Argentina, 6, 41

\bibitem[{{Simien} \& {Prugniel}(2002)}]{Simien2002}
{Simien}, F. \& {Prugniel}, P. 2002, \aap, 384, 371

\bibitem[{{Smith} {et~al.}(2008){Smith}, {Marzke}, {Hornschemeier}, {Bridges},
  {Hudson}, {Miller}, {Lucey}, {V{\'a}zquez}, \& {Carter}}]{Smith2008}
{Smith}, R.~J., {Marzke}, R.~O., {Hornschemeier}, A.~E., {et~al.} 2008, \mnras,
  386, L96

\bibitem[{{Smith Castelli} {et~al.}(2008){Smith Castelli}, {Bassino},
  {Richtler}, {Cellone}, {Aruta}, \& {Infante}}]{SmithCastelli2008}
{Smith Castelli}, A.~V., {Bassino}, L.~P., {Richtler}, T., {et~al.} 2008,
  \mnras, 386, 2311

\bibitem[{{Thuan}(1985)}]{thu85}
{Thuan}, T.~X. 1985, \apj, 299, 881

\bibitem[{{Toloba} {et~al.}(2009){Toloba}, {Boselli}, {Gorgas}, {Peletier},
  {Cenarro}, {Gadotti}, {Gil de Paz}, {Pedraz}, \& {Yildiz}}]{Toloba2009}
{Toloba}, E., {Boselli}, A., {Gorgas}, J., {et~al.} 2009, \apjl, submitted

\bibitem[{{Tonnesen} {et~al.}(2007){Tonnesen}, {Bryan}, \& {van
  Gorkom}}]{Tonnesen2007}
{Tonnesen}, S., {Bryan}, G.~L., \& {van Gorkom}, J.~H. 2007, \apj, 671, 1434

\bibitem[{{Trentham} \& {Tully}(2002)}]{Trentham2002}
{Trentham}, N. \& {Tully}, R.~B. 2002, \mnras, 335, 712

\bibitem[{{Tully} \& {Fisher}(1977)}]{TullyFisher}
{Tully}, R.~B. \& {Fisher}, J.~R. 1977, \aap, 54, 661

\bibitem[{{Tully} \& {Trentham}(2008)}]{Tully2008}
{Tully}, R.~B. \& {Trentham}, N. 2008, \aj, 135, 1488

\bibitem[{{Vaduvescu} \& {McCall}(2008)}]{Vaduvescu2008}
{Vaduvescu}, O. \& {McCall}, M.~L. 2008, \aap, 487, 147

\bibitem[{{Vaduvescu} {et~al.}(2006){Vaduvescu}, {Richer}, \&
  {McCall}}]{Vaduvescu2006}
{Vaduvescu}, O., {Richer}, M.~G., \& {McCall}, M.~L. 2006, \aj, 131, 1318

\bibitem[{{Valcke} {et~al.}(2008){Valcke}, {De Rijcke}, \&
  {Dejonghe}}]{Valcke2008}
{Valcke}, S., {De Rijcke}, S., \& {Dejonghe}, H. 2008, \mnras, 389, 1111

\bibitem[{{van den Bergh}(1986)}]{vandenBergh1986}
{van den Bergh}, S. 1986, \aj, 91, 271

\bibitem[{{van Zee} {et~al.}(2004{\natexlab{b}}){van Zee}, {Barton}, \&
  {Skillman}}]{vanZee2004b}
{van Zee}, L., {Barton}, E.~J., \& {Skillman}, E.~D. 2004{\natexlab{b}}, \aj,
  128, 2797

\bibitem[{{van Zee} {et~al.}(2004{\natexlab{a}}){van Zee}, {Skillman}, \&
  {Haynes}}]{vanZee2004a}
{van Zee}, L., {Skillman}, E.~D., \& {Haynes}, M.~P. 2004{\natexlab{a}}, \aj,
  128, 121

\bibitem[{{Whitmore} {et~al.}(1993){Whitmore}, {Gilmore}, \&
  {Jones}}]{Whitmore1993}
{Whitmore}, B.~C., {Gilmore}, D.~M., \& {Jones}, C. 1993, \apj, 407, 489

\bibitem[{{Yoshii} \& {Arimoto}(1987)}]{YoshiiArimoto1987}
{Yoshii}, Y. \& {Arimoto}, N. 1987, \aap, 188, 13

\bibitem[{{Young} \& {Currie}(1994)}]{Young1994}
{Young}, C.~K. \& {Currie}, M.~J. 1994, \mnras, 268, L11

\end{thebibliography}


\end{document}